# Influence of microwave annealing on GMI response and magnetization of an amorphous $Fe_{73.5}Nb_3Cu_1Si_{13.5}B_9$ ribbon.

B. Kaviraj[1] and S.K Ghatak

*Department of Physics and Meteorology, Indian Institute of Technology, Kharagpur, India.*

**Abstract**

*The resistive and reactive components of magneto-impedance was studied for the as-quenched and microwave annealed amorphous $Fe_{73.5}Nb_3Cu_1Si_{13.5}B_9$ ribbon as a function of biasing d.c magnetic fields (-60 to +60 Oe) and excitation frequencies (0.1, 1, 10 and 20MHz). The magneto-impedance (both components) response was much reduced for the microwave annealed ribbon and the changes were more discernable at higher excitation frequencies. The imaginary component of magneto-impedance showed maxima at finite (non-zero) d.c magnetic fields for both the as-quenched and microwave annealed ribbons. Magnetization measurements performed for both the as-quenched and microwave annealed ribbons revealed the magnetic hardness of the latter. The initial susceptibility decreases by two orders of magnitude for the microwave-annealed ribbon. XRD measurements indicated the transformation of the surface of the ribbon from the amorphous state to the crystalline one.*

---

[1] Electronic Address: bhaskar@phy.iitkgp.ernet.in

# Introduction

Giant Magneto-impedance (GMI) in amorphous wires and ribbons has attracted much attention in the last few years because of its prospective application in magnetic recording heads and sensor elements [1]. The magneto-impedance effect consists of a large variation in the impedance (both the resistive and reactive components) of a soft magnetic material submitted to a static magnetic field. This effect has been observed in different types of samples: amorphous wires and microwires, ribbons, thin films, and also in nanocrystalline wires and ribbons [2-9].

It is generally understood that the combined effect of electromagnetic screening and the magnetization dynamics in a ferromagnetic conductor leads to the GMI effect. In the domain of low frequency the MI effect is considered as a classical phenomenon and can be explained by solving the Maxwell equations coupled with magnetization dynamics. The impedance of a conductor is governed by the length scales: thickness of ribbon and field penetration depth. The penetration depth, $\delta$ is given by $\delta \sim (\sigma \mu f)^{-1/2}$, where $\sigma$ is the electrical conductivity, f is the drive current frequency and $\mu$ is the magnetic permeability due to a.c. magnetic field. When $\delta$ is larger than thickness, the sample provides low impedance path , and the opposite situation creates high impedance for current flow. So, the impedance for a given frequency can be modified by a change in the permeability. In soft ferromagnetic material the magnetic response as measured by the permeability depends on number of factors e.g. the magnetic anisotropy, biasing field, the frequency and amplitude of a.c magnetic field. The magnetic anisotropy in amorphous ferromagnetic state is dominated by interaction between frozen stress and magnetization. The magnetic anisotropy thereby can be altered by adequate heat treatments and it has been borne out by experiments [10].

As quenched Fe based alloys with nominal composition $Fe_{73.5}Nb_3Cu_1Si_{13.5}B_9$ exhibits magneto-impedance change upto 90% [11]. Knobel et. al annealed Fe-Cu-Nb-Si-B series of alloy at 550-600$^0$C and obtained maximum value of MI ratio of the order of 100% [12,13]. The r.f. current flows within the skin depth, and any modification of magnetic structure within layers of skin depth can therefore impede the current flow. Such modification can be done by treating sample in presence of high amplitude microwave radiation and consequently MI would be changed. Motivated with this fact we studied the influence of microwave annealing on the magneto-impedance and magnetic properties of an amorphous $Fe_{73.5}Nb_3Cu_1Si_{13.5}B_9$ ribbon and observed a large change in MI and its magnetic field dependence.

**Experimental**

The experiments were carried out with samples cut from amorphous $Fe_{73.5}Nb_3Cu_1Si_{13.5}B_9$ ribbon prepared using conventional single roller rapid quenching in vaccum. The samples 30µm thick and 4mm wide was cut into 20mm length and is placed within a small secondary coil of rectangular geometry with 50 turns and length of the sample was larger than that of the secondary coil. The real and the imaginary component of the impedance Z=R+jX, where X=ωL (L being the inductance of the sample) were obtained using a Agilent Impedance Analyzer (Model-4294A). The sample was connected to the analyzer with the accessory 16048B test lead, which contained two coaxial cables. The cables were 100 cm long and permitted the ribbon to sit within a Helmholtz coil of sensitivity 16Oe/A, which produced a magnetic field -60≤ H ≤ 60 Oe. The applied d.c field was parallel to the a.c field produced by the current carrying coil wounded across the sample. The drive current amplitude was fixed at 20mA. This corresponded to an a.c excitation field of about 140A/m. The ribbon axis was normal to the Earth's magnetic field. The values R and X reported here refers to the impedance in series combination. The real and imaginary parts of the MI ratio are represented as:

$$\left(\frac{\Delta R}{R}\right)\% = \frac{[(R(H_{dc}) - R(H=0))]}{R(H=0)}\%$$

$$\left(\frac{\Delta X}{X}\right)\% = \frac{[(X(H_{dc}) - X(H=0))]}{X(H=0)}\%$$

The as-quenched ribbons were then annealed in a microwave atmosphere for about 10min. and then the magneto-impedance measurements were performed on them.

Magnetization measurements were performed on both the as-quenched and microwave annealed samples by fluxmetric method at room temperature.

## Results and Discussions

In Fig 1(a) and 1(b), the resistive and the reactive parts of the magneto-impedance is plotted as a function of external d.c magnetic field H at an excitation frequency of 100KHz for both the as-quenched and microwave annealed samples.

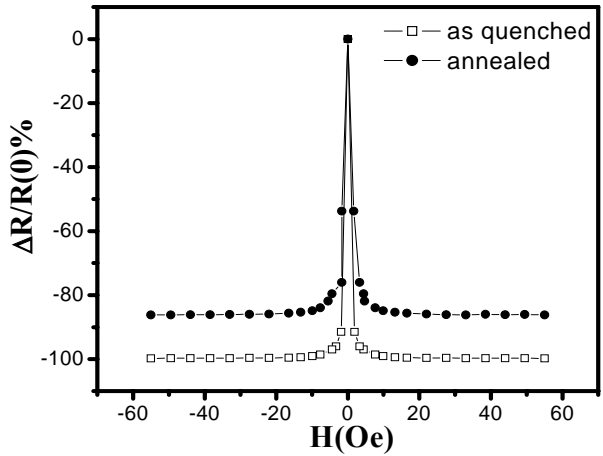 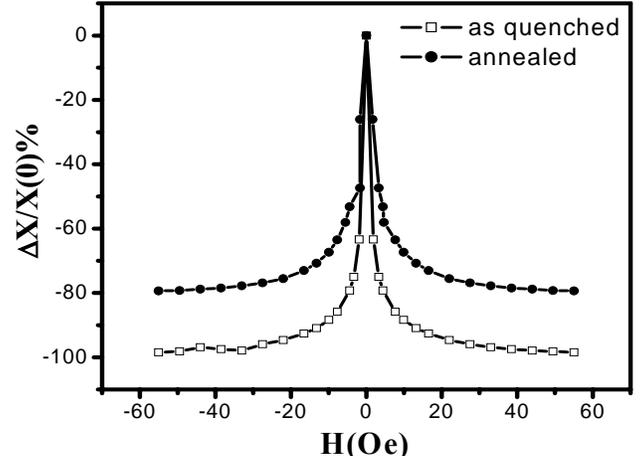

**Fig.1.** Relative resistance (ΔR/R)% (a) and (ΔX/X)% (b) as a function of H for ω =0.1MHz for the as-quenched (open box) and microwave-annealed (solid circle) $Fe_{73.5}Nb_3Cu_1Si_{13.5}B_9$ ribbon. The applied d.c field was in the plane of the ribbon and parallel to the exciting a.c field.

For small d.c field, field dependence of MI is similar for both samples, however the decrease in the resistive part is much sharper compared to reactive part. The as-quenched samples exhibited 99% changes in both the real and imaginary parts of MI, whereas for the microwave annealed samples, the change in the real part of MI% (85%) was larger than that of the imaginary (nearly 80%). Such high changes in real and imaginary parts of magneto-impedance can be attributed to the decrease of transverse permeability in the presence of biasing d.c field.

Fig 2(a) and 2(b) shows the MI response of the as-quenched and microwave-annealed samples at a higher frequency of 1MHz.

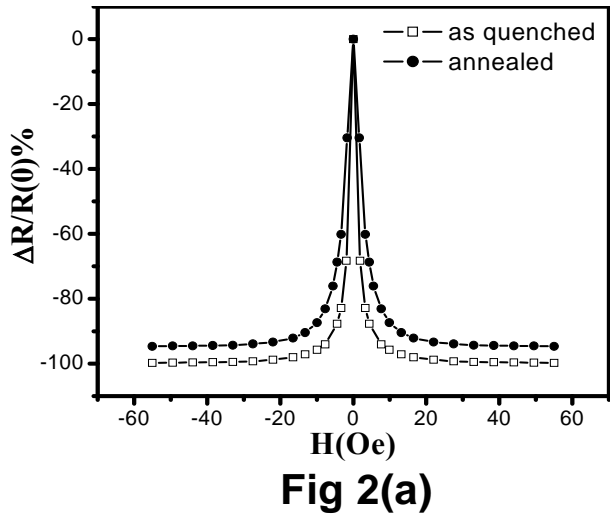 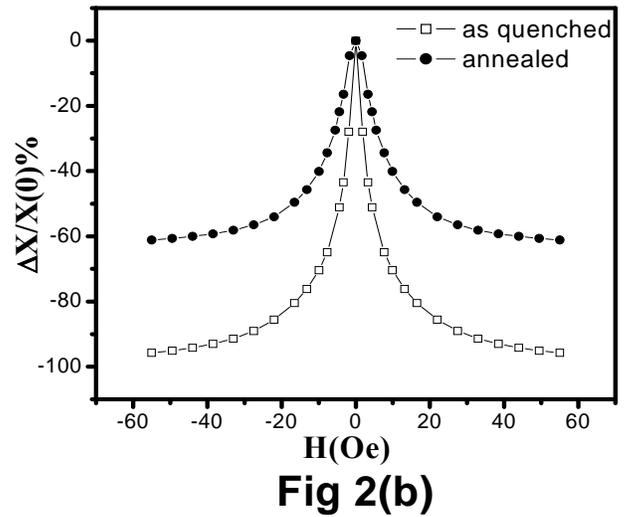

**Fig.2.** Relative resistance (ΔR/R)% (a) and (ΔX/X)% (b) as a function of H for ω =1MHz.

There is an appreciable change in the reactive component of MI% for the sample undergoing microwave treatment compared to as-quenched one. The maximum MI, which was nearly 95% for the as quenched ribbon, decreases to 60% after exposure to the microwave radiation. Similar to earlier case, the field variation of the resistive part of MI% is sharper than that of the reactive part.

With the increase in excitation frequency, both components of MI are reduced, and the results are shown for 10MHz (Fig 3) and 20MHz (Fig 4).

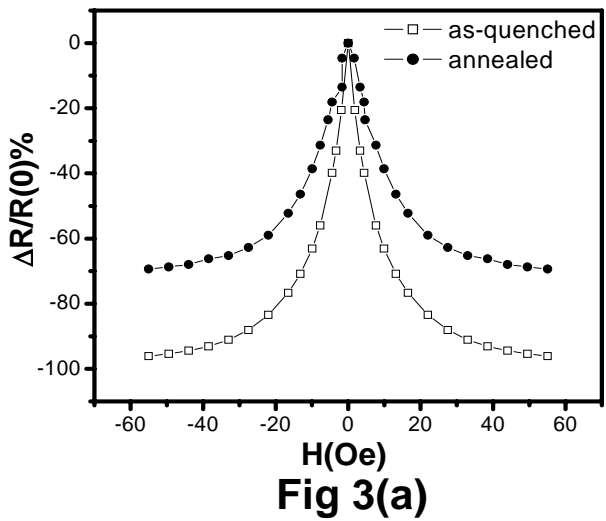
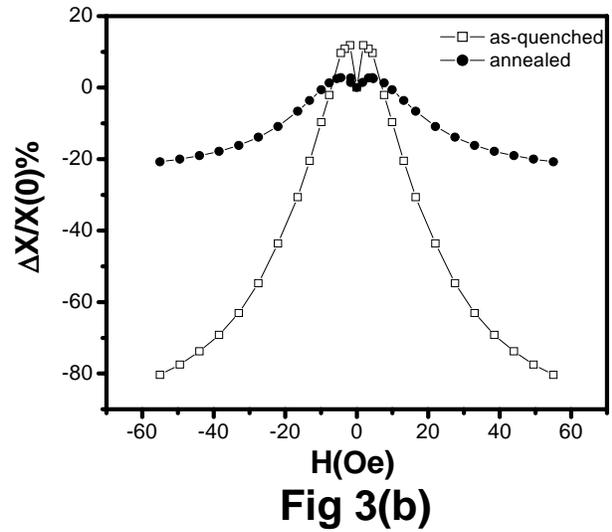

**Fig.3.** Relative resistance (ΔR/R)% (a) and (ΔX/X)% (b) as a function of H for ω =10MHz.

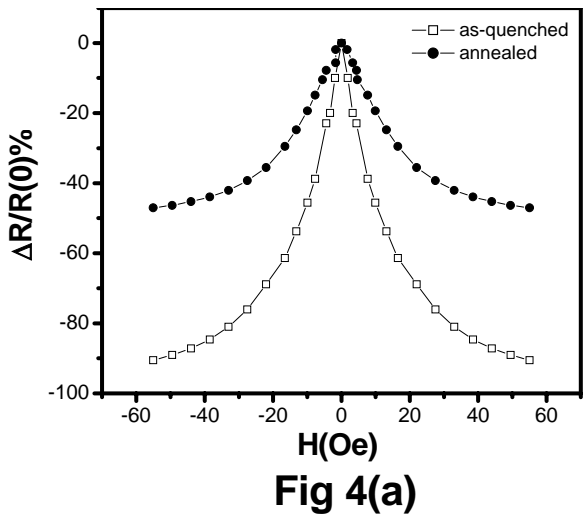
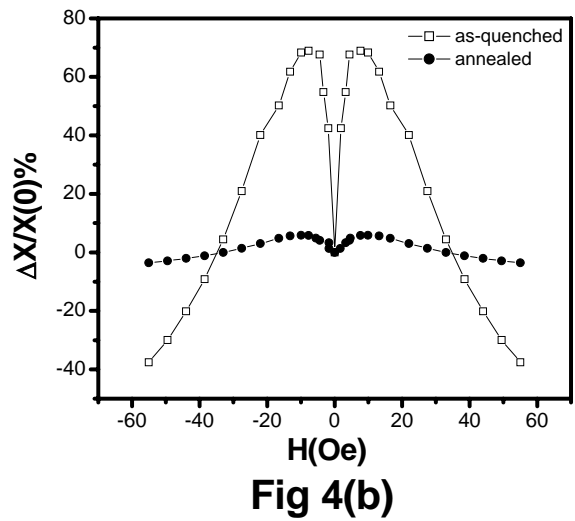

**Fig.4.** Relative resistance (ΔR/R)% (a) and (ΔX/X)% (b) as a function of H for ω =20MHz.

In this frequency range the reactive parts of MI for the as-quenched and the microwave annealed sample differ significantly. The maximum change of the reactive part of MI for the field annealed sample decreases to only 20% and 5% at 10 and 20 MHz respectively at maximum 60Oe d.c field whereas for the as-quenched sample the corresponding values are 80% and 40% at same field. The reactive part $\Delta X$ becomes positive at low field and exhibits peak at finite field. However, resistive part decreases with magnetic field. As the magnitude of reactive part is higher than that of resistive part, the MI increases with magnetic field at low H. The peak value of $(\Delta X/X)\%$ for the as-quenched ribbon strongly depends on frequency and it changes from 12% at 10MHz to 70% at 20MHz and the peak position shifts to higher field with frequency. In annealed sample the sensitivity of peak value is drastically reduced, but the position of the peak appears at higher field. For the as-quenched sample, the peak-field, $H_k \sim 1.8$Oe was lower than that of the microwave annealed sample with $H_k \sim 3.4$Oe at 10MHz and corresponding values of peak field at 20MHz were $H_k \sim 4.8$Oe and 7.4Oe for the as-quenched and microwave-annealed samples respectively. We note that the field required to saturate the reactive part of MI was much lower for the microwave-annealed sample as compared to the as-quenched sample where fields greater than 60Oe are required to saturate the impedance. For the resistive part of MI at 10 MHz, the maximum value for the annealed sample was 70%, which is much lower compared to 95% of the as-quenched sample. The maximum change in resistance (%) at 20MHz is again reduced to 40% in annealed sample. Comparing the microwave responses within these frequency regions it is clear that the difference in maximum MI ratios between the as-quenched and the microwave-annealed sample increases with increase in frequency. The difference is much more prominent in the reactive part of MI. This is more prominent at 20MHz (Fig.4).

The magnetization measurements have been performed for both the samples at room temperature using an a.c magnetometer using magnetizing field of amplitude of 400G at 70Hz. Fig 5 depicts the reduced initial magnetization curves of as-quenched and microwave-annealed samples.

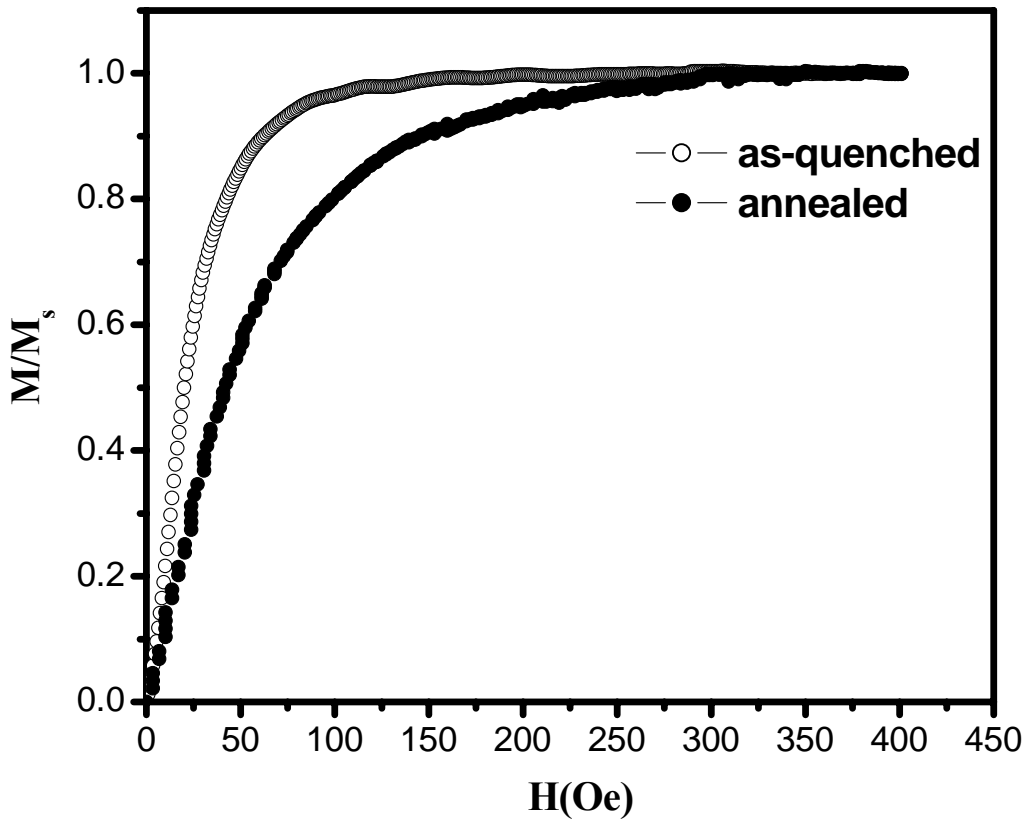

**Fig (5)**

**Fig 5 – Initial magnetization curve of sample in as-quenched and annealed state.**

It is evident that the initial magnetization process is slowed down in treated sample. The initial magnetization of treated sample is nearly reduced by 30 times compared to that of as-quenched one. As-quenched sample attains technical saturation at field around 100G, whereas a higher field is required for saturating annealed sample. The changes in the magnetic parameters (as observed in the magnetization measurements) of the amorphous $Fe_{73.5}Nb_3Cu_1Si_{13.5}B_9$ ribbon upon microwave annealing indicates that the sample became magnetically harder upon exposure to microwave radiation. This is due to increase in magnetic anisotropy energy resulting from microwave absorption. In amorphous state the anisotropy is mainly magnetostrictive in origin and small as is borne out from softness of magnetization process. Microwave is absorbed within few atomic layers due to very small electric field penetration and thereby the atomic arrangement within these exposed layers are strongly modified. In order to get an idea of such atomic rearrangement low angle x-ray scattering spectra has been taken with spectrometer (Phillips X'Pert Pro) and they are presented in Fig 6(a) and 6(b).

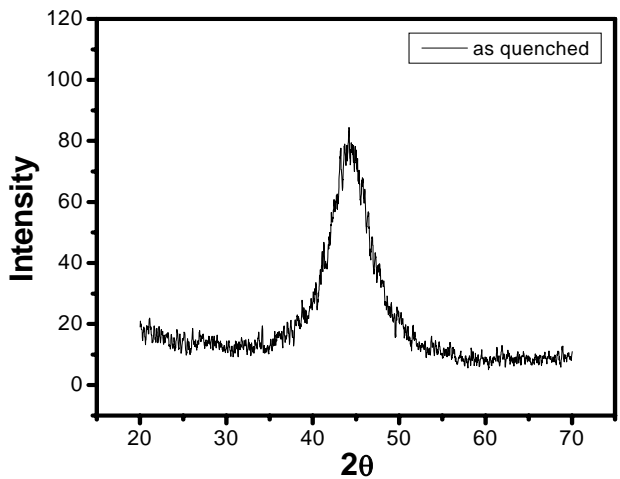
Fig 6(a)

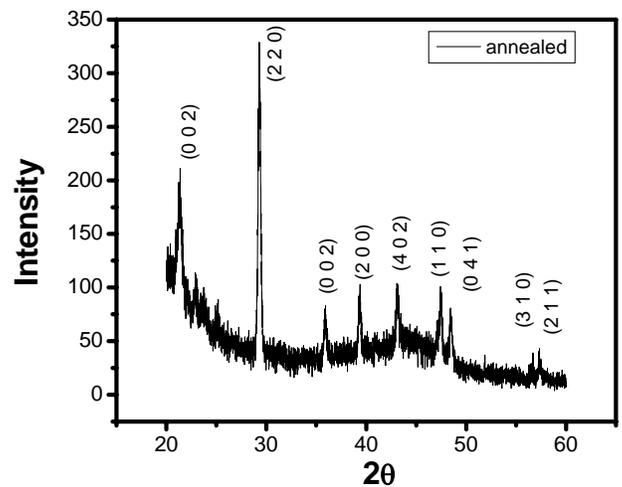
Fig 6(b)

**Fig 6: XRD patterns of the $Fe_{73.5}Nb_3Cu_1Si_{13.5}B_9$ ribbon in the as quenched (a) state and after microwave annealing (b).**

The XRD patterns as obtained with the incident (Cu-K$_\alpha$) radiation is at $3^0$ grazing angle. The spectra of the as-cast and annealed ribbon clearly indicate the structural modifications after microwave treatment. Considering small penetration depth of radiation the devitrification occurs within few layers below the surface and hence spectra provide information about changes in atomic environment near the surface. The peaks in spectra are result of growth of microcrystalline phases and are identified with the crystalline phases of $FeSi_2$, $Fe_2Si$, $Fe_5Si_3$ and FeB.

The impedance of the ribbon of thickness 2d (30μm for the sample) excited with a.c field parallel to length of the ribbon and with frequency ω can be obtained from the solution of Maxwell equations and magnetization equation and is given by

$$Z = -jX_0\mu\left(\frac{\tanh(kd)}{kd}\right) \qquad (1)$$

where $X_0$ = reactance of the signal coil without the sample, μ = effective permeability of the sample due to excitation. The propagation vector is given by $k = (1+j)/\delta$, where effective skin depth, $\delta = [\omega \sigma \mu \mu_0 /2]^{-1/2}$, σ being the conductivity. The conductivity of the ribbon at room temperature is $3.2 \times 10^5/\Omega$-m and the permeability in this alloy is very high and lies in the range of $10^4 - 10^5$ [14]. Taking $\mu \approx 5 \times 10^4$ at zero bias field which is close to slope of initial magnetization curve, the skin depth δ is changing from 12 μm at 0.1MHz to 1.2 μm at 10MHz. Smaller value of skin depth is mainly due to larger magnetic response of the sample. With microwave treatment the surface is partly devitrified and this microcrystalline state near surface layers has a higher magnetic anisotropy resulting smaller value of permeability and higher value of skin depth. The variation of permeability with biasing field is also diminished due to higher anisotropy field, and this in turn results smaller variation of both components of impedance. In the annealed sample, the initial magnetization process (Fig.6) is observed to be nearly 30 times slower. At high frequency the field variation of the skin depth is much reduced and this makes impedance less susceptible to field. Assuming extreme situation of d/ δ, it follows from eqn.1:

$$\left(\frac{\Delta X}{X}\right)\% = \frac{\left[(\mu_r + \mu_i)_H - (\mu_r + \mu_i)_{H=0}\right]}{(\mu_r + \mu_i)_{H=0}}$$

$$\left(\frac{\Delta R}{R}\right)\% = \frac{\left[(\mu_r - \mu_i)_H - (\mu_r - \mu_i)_{H=0}\right]}{(\mu_r - \mu_i)_{H=0}}$$

where $\mu_r$ ( $\mu_i$ ) is the in-phase (out-of-phase) part of permeability. In general $\mu_i < \mu_r$ and difference decreases as frequency tends towards the relaxation frequency. In annealed sample, $\mu_r$ and its field dependence at higher frequency decreases due to increased anisotropy field with devitrification. At high frequency $\mu_i$ is appreciable and this leads to lower value of (ΔX/X)% compared to that of (ΔR/R)%.

## Conclusions

The magneto-impedance and magnetization measurements were carried out for both the as-quenched and microwave-annealed ribbons. Appreciable decrease of magneto-impedance in microwave-annealed ribbons especially at higher frequencies was due to the induced anisotropy by the formation of microcrystallines upon exposure of microwave. These results demonstrate that the GMI phenomenon is the combined effect of screening of e.m field and higher magnetic response of ferromagnetic metallic system. Magnetization and susceptibility measurements indicated that the material indeed has grown harder magnetically. XRD patterns showed the change in surface properties, which were responsible for the change in MI behaviour of the ribbon before and after the microwave treatment.